**Mechanical, optoelectronic and thermoelectric properties of half-Heusler p-type semiconductor BaAgP: A DFT investigation**

F. Parvin[a], M. A. Hossain[b]\*, M. I. Ahmed[a], K. Akter[a] & A.K.M.A. Islam[a,c]

[a]*Department of Physics, University of Rajshahi, Rajshahi-6205, Bangladesh*
[b]*Department of Physics, Mawlana Bhashani Science and Technology University, Santosh, Tangail-1902, Bangladesh;*
[c]*International Islamic University Chittagong, Kumira, Chittagong-4318, Bangladesh*

Email\*: anwar647@mbstu.ac.bd

**Abstract**

We have explored the mechanical, electronic, optical and thermoelectric properties of p-type half-Heusler compound BaAgP for the first time using density functional theory based calculations. The mechanical and dynamical stability of this compound is confirmed by studying the Born stability criteria and phonon dispersion curve, respectively. It is soft, ductile and elastically anisotropic. The atomic bonding along *a*-axis is stronger than that along *c*-axis. The calculated electronic structure reveals that the studied compound is an indirect band gap semiconductor. The analysis of charge density distribution map and Mulliken population reveals that the bonding in BaAgP is a mixture of covalent and ionic. The optical features confirm that BaAgP is optically anisotropic. The high absorption coefficient and low reflectivity in the visible to ultraviolet region make this compound a possible candidate for solar cell and optoelectronic device applications. The thermoelectric properties have been evaluated by solving the Boltzmann semi-classical transport equations. The calculated power factor at 1000K along a-axis is 35.2 $\mu W/cmK^2$ (with $\tau=10^{-14}$ s) which is ~3.5 times larger than that of SnSe, a promising layered thermoelectric materials. The thermoelectric figure of merit, ZT of BaAgP is 0.44 which is small due to high thermal conductivity. So the reduction of thermal conductivity is essential to enhance thermoelectric performance of BaAgP in device applications.

**Keywords:** Mechanical; Optoelectronic; Thermoelectric; Half-heusler; Density functional theory

## 1. Introduction

At present, two of the major problems in our society are fuel shortages and environmental pollution. These problems arose due to the huge dependence on fossil fuels. To solve these problems it is especially important to exploit new energy conversion techniques which have high efficiency, and are also green and eco-friendly [1]. Thermoelectric materials, which can generate electricity from waste heat or be used as solid-state Peltier coolers, could play an important role in a global sustainable energy solution [2]. Nowadays the half-Heusler (HH) compounds have been attracted intense attention in the field of the condensed matter physics community due to their desirable characteristics, such as excellent mechanical robustness, good thermoelectric performance, high-temperature stability and low toxicity [3]. These compounds have potential application in spintronics (spin transport of electrons) and the green energy-related fields, such as solar cells or thermoelectrics (TE) [4]. Generally, the half-metallic materials are used for spintronic device and semiconducting materials are used for solar cell and thermoelectric devices. The potentiality of a Heusler compound for a particular type of application can be easily determined by counting their valence electrons [5]. The HH materials with 18 valence electrons count (VEC) in the unit cell show thermoelectric properties. Many members of this family such as ScRhTe [6] TiCoSb [5], ScPtSb [7], TiNiSn [8], LiMgN [9], ScNiGa [10], PdZrGe [11], RhTiSb [12], RhFeX (X=Sn, Ge) [13], TaCoSn [14], HfNiSn [15], XNiSn [16] are considered as thermoelectric materials due to their small thermal conductivity, high Seebeck coefficient and *ZT* value close to unity. Again, the compounds with low reflection and minimal light loss, and which have greatest absorption in the visible to ultraviolet region, are considered as most appropriate for solar cell and optoelectronic devices [17].

BaAgP, a HH compound, has already been synthesized [18] and its lattice parameters are determined. It has hexagonal crystal structure with space group of $P6_3/mmc$ - $D_{6h}^4$ (No. 194). Although, BaAgP and BaAgAs are isostructural compounds they show different behavior. Recently, S. Mardanya *et. al.* [19] calculated the electronic structure of $BaAgAs_{1-x}P_x$ using DFT with PAW pseudopotentials [20]. Their analysis shows that the end compounds BaAgAs and BaAgP are semimetal and trivial insulator, respectively. To the best of our knowledge there are no other available experimental or theoretical studies on the mechanical, electronic, optical, vibrational, and thermoelectric properties in the literature. All of these properties are very important for technological applications. Elastic constants and elastic moduli determine the mechanical response of a compound under external stress. The knowledge regarding electronic properties and frequency dependent optical constants are essential for optoelectronic applications. The thermoelectric response of a material can be understood by its dimensionless figure of merit ($ZT = S^2\sigma T/\kappa$, where S, $\sigma$, $\kappa$ and T stands for Seebeck coefficient, electrical conductivity, thermal conductivity consisting of electronic and lattice contribution and absolute temperature, respectively) which largely depends on Seebeck coefficient, electrical conductivity and thermal conductivity of that material.

In this research paper we have conducted a detailed and systematic study on the structural, elastic, electronic, optical and thermoelectric properties of BaAgP. To analyze the dynamical stabilities of this compound we have also calculated phonon properties such as phonon dispersion curve and phonon density of states.

## 2. Computational details

The crystal structure of BaAgP has been optimized using CASTEP code [21] which is based on the density functional theory (DFT) [22, 23]. The plane wave basis set cut-off is set as 500 eV, and for the sampling of the Brillouin zone [24], a $19 \times 19 \times 8$ Monkhorst–Pack mesh is employed [25]. The convergence thresholds of $5 \times 10^{-6}$ eV/atom for the total energy, 0.01 eV/Å for the maximum force, 0.02 GPa for the maximum stress and $5 \times 10^{-4}$ Å for maximum displacement were used to achieve geometry optimization. This optimized crystal structures have been used for all calculations of the present study. The IRelast method [26] interfaced with the full potential linearized augmented plane wave (LAPW) as implemented in WIEN2k [27] was employed for the calculation of elastic properties. The generalized gradient approximation (GGA) within the Perdew Burke-Ernzerhof (PBE) [23, 28] scheme was utilized for elastic properties calculation. The electronic band structures were calculated in WIEN2k using GGA-PBE and Tran-Blaha modified Becke-Johnson potential (TB-mBJ) [29]. To obtain a good convergence basis set, a plane wave cut off of kinetic energy RKmax = 7.0 was selected by convergence test. The optical and transport properties were calculated using a dense mesh ($35 \times 35 \times 15$) k-points. The TB-mBJ potential was used for optical and thermoelectric transport properties calculations. The Boltzmann semi-classical transport equations as implemented in BoltzTraP code [30] were employed to calculate transport properties at constant relaxation time $\tau$. The standard equations of transport coefficients are defined in the Boltzmann transport theory are found elsewhere [31-33].

## 3. Results and discussion

### 3.1 Structural and Mechanical properties

The ternary compounds, BaAgP and BaCuP are isotypic and crystallizes in a modified $Ni_2In$ - structure (hexagonal system) [18]. The unit cell of BaAgP compound is shown in Fig. 1. The Ba, Ag and P atoms are at special sites of: 2a (0, 0, 0) for Ba, 2d (1/3, 2/3, 3/4) for Ag and 2c (1/3, 2/3, 1/4) for P. The optimized structural parameters are presented in Table 1 along with experimental result [18] for comparison. It is seen that our estimated lattice parameters fairly agree with the experimental value.

**Table 1.** Calculated lattice parameters (in Å), unit cell volume (in Å$^3$) and single crystal elastic constants $C_{ij}$ (in GPa) of BaAgP along with the experimental data.

| $a$ | $c$ | $V$ | $C_{11}$ | $C_{12}$ | $C_{13}$ | $C_{33}$ | $C_{44}$ | $C_{66}$ | Ref. |
|-----|-----|-----|----------|----------|----------|----------|----------|----------|------|
| 4.505 | 9.013 | 158.41 | 115.8 | 31.1 | 22.8 | 71.3 | 30.4 | 42.3 | This |
| 4.496 | 8.828 | - | - | - | - | - | - | - | [1]$^{Expt.}$ |

The elastic stiffness constants is important for the information about the stability of structure of any semiconductor. In the case of hexagonal structure, there are six independent elastic stiffness constants $C_{ij}$, i.e., $C_{11}$, $C_{12}$, $C_{13}$, $C_{33}$, $C_{44}$ and $C_{66}$. Our calculated elastic constants completely satisfy the Born stability criteria [34], ($C_{11} > 0$, $C_{11}-C_{12} > 0$, $C_{44} > 0$, $(C_{11}+C_{12})C_{33}-2C^2_{13} > 0$), which indicate that BaAgP is mechanically stable.

The elastic constants $C_{11}$ and $C_{33}$ have a perpendicular slope which indicates that there is a very strong resistance along $a$ and $c$ axes, respectively. It is obvious from Table 1 that $C_{11} > C_{33}$ so the atomic bonding is stronger along $a$-axis than that along $c$-axis. Since the elastic constants $C_{11}$ and $C_{33}$ are larger than $C_{44}$, the linear compression along the crystallographic $a$- and $c$-axis is rather difficult in comparison with the shear deformation.

**Table 2.** Calculated bulk modulus, $B$ (GPa), shear modulus, $G$ (GPa), Young's modulus, $Y$ (GPa), Pugh's ratio, $B/G$, Poisson's ratio, $v$, Zener anisotropy factor $A$ of BaAgP.

| $B$ | $G$ | $Y$ | $B/G$ | $v$ | $A$ | Ref. |
|-----|-----|-----|-------|-----|-----|------|
| 49.2 | 35.0 | 85.0 | 1.41 | 0.21 | 0.717 | This calc. |

The calculated polycrystalline elastic parameters $B$, $G$, $Y$ are derived from single crystal elastic constants using the so-called Voigt-Reuss-Hill formalisms [35-37] and are listed in Table 2. Smaller value of $G$ compared to $B$ (Table 2) indicates that the mechanical stability will be dominated by shear modulus. The Young's modulus, which is the ratio between tensile stress and strain, is the measurement of the resistance (stiffness) of an elastic solid to a change in its length [38, 39] and provides with a measure of thermal shock resistance.

The elastic constant is also related to the ductile or brittle properties of the materials by the empirical relation $B/G$ [40, 41], where shear modulus $G$ represents plastic deformation and bulk modulus $B$ shows the resistance to fracture [42]. If $B/G > 1.75$, the material has ductile nature, otherwise behaves in a brittle manner [14]. In the present calculation, the value of $B/G$ has been found to be 1.41, which indicates that BaAgP is brittle in nature.

The value of Poisson's ratio $v$ provides the information about the characteristics of the bonding forces. The $v = 0.25$ and 0.5 are the lower and upper limit for the central forces in solids, respectively [43]. Our calculated value of $v$ is equal to 0.21, which is less than 0.25. This indicates that BaAgP is affected by a certain amount of non-central forces revealing the presence of covalent character in the bonding.

Elastic anisotropy is an important design parameter, especially for layered structured compounds. It has significant implications in engineering science due to its correlation with the possibility of creation and propagation of microcracks in the crystals [44]. Zener anisotropy factor ($A$) have been calculated using the relation [45] $A = 2C_{44}/(C_{11}-C_{12})$ and found to be 0.717. This shows that BaAgP is an anisotropic material. The linear compressibility along the $a$-axis ($\chi_a$) and $c$-axis ($\chi_c$) can be calculated using the following relations [46]:

$$\chi_a = -\frac{1}{a}\frac{\partial a}{\partial P} = \frac{C_{33} - C_{13}}{C_{33}(C_{11} + C_{12}) - 2C_{13}^2}$$

$$\chi_c = -\frac{1}{c}\frac{\partial c}{\partial P} = \frac{C_{11} + C_{12} - 2C_{13}}{C_{33}(C_{11} + C_{12}) - 2C_{13}^2}$$

The calculated values of $\chi_a$ and $\chi_c$ are $5.14 \times 10^{-3}$ GPa$^{-1}$ and $10.74 \times 10^{-3}$ GPa$^{-1}$, respectively. These values indicate that the compressibility along $c$-axis is almost double than that along $a$-axis. Debye temperature ($\Theta_D$) is a characteristic parameter of crystals. A number of important physical properties such as melting temperature, phonon specific heat, lattice thermal conductivity etc. of solids depend on Debye temperature. In this study we have calculated the Debye temperature using mean sound velocities [47] as $\Theta_D = \frac{h}{k_B} \left[ \frac{3n}{4\pi V} \right]^{1/3} v_m$, where $n$ is the no. of atoms in the unit cell, $V$ is the volume of the unit cell and $v_m$ is the mean sound velocity. The mean sound velocity in the crystal can be calculated using the relation $v_m = \left[ \frac{1}{3} \left( \frac{1}{v_l^3} + \frac{2}{v_t^3} \right) \right]^{-1/3}$, here $v_t$ and $v_l$ are the transverse and longitudinal sound velocities, respectively in an isotropic crystal. These can be determined with the help of bulk modulus, $B$ and shear modulus, $G$ using the equations $v_l = \left[ \frac{3B+4G}{3\rho} \right]^{1/2}$ and $v_t = \left[ \frac{G}{\rho} \right]^{1/2}$.

**Table 3.** Calculated density ($\rho$), longitudinal, transverse, mean sound velocities ($v_l$, $v_t$ and $v_m$), and Debye temperature ($\Theta_D$), of BaAgP.

| $\rho$ (gm/cm$^3$) | $v_l$ (km/s) | $v_t$ (km/s) | $v_m$ (km/s) | $\Theta_D$ (K) | Ref. |
|---|---|---|---|---|---|
| 5.79 | 4.07 | 2.46 | 2.72 | 272 | This calc. |

The calculated Debye temperature is low (272 K) and this lower Debye temperature indicates the possibility of lower lattice thermal conductivity.

## 3.2 Electronic Properties

### 3.2.1 Band structure and Density of states

The calculation of the electronic band structure is important to understand the physical properties of crystalline solids which almost completely explain optical as well as transport properties. We have calculated the electronic band structure for BaAgP along high symmetry directions within the *k*-space and is displayed in Fig. 2. The GGA-PBE potential underestimate the experimental band gap by around 50% [48] and a lot of studies are available in literature that TB-mBJ potential can predict band gap comparable with experimental values [49-54]. In this study, we have calculated the electronic band structures using two different exchange-correlation functionals GGA-PBE and TB-mBJ. From these figures it is noted that the top of the valence band is at $\Gamma$-point but the bottom of the conduction band is at *M*-point. Hence the studied compound is an indirect band gap semiconductor and the calculated band gaps are 0.15 and 0.59 eV for GGA-PBE and TB-mBJ functional, respectively. The Fermi level $E_F$ is shown as broken line (Fig. 2). The contribution of different atomic states to the band structure can be further elucidated from the (DOS) calculation. Generally, the density of states (DOS) is defined as the number of available electronic states per unit energy per unit volume. We need to know the density of states in order to calculate various optical properties and how electrons and holes distribute themselves within a solid. The total and partial densities of states for BaAgP are presented in Fig. 3 for TB-mBJ and PBE functionals. From Fig. 3, it is seen that the lowest part of valence band within -4 eV to ~ -3 eV is comprised of Ag-4*d*, Ag-5*s* and P-3*p* states. The next region in the valence band within the energy range of -3 eV to 0 eV is mainly originates by the hybridization of Ag-4*d* and P-3*p* states with small contribution of Ag-4*p* states. An energy band gap exists at the Fermi level. The higher energy band above the Fermi energy is formed mainly by the Ba-4*d* states with small contributions of Ag-4*p*, 4*d* and 5*s* electronic orbitals.

### 3.2.2 Charge density and Mulliken population analysis

To explain the transfer of charge and to visualize the bonding characteristics of BaAgP, the electronic charge density distribution mapping along (101) of BaAgP is presented in Fig. 4. The accumulation of charges (positive regions) between two atoms constitutes the covalent bonds. On the other hand the balancing of positive or negative (depletion regions) charge at the atomic position signifies ionic bonding [55]. The accumulation of charges can be found in Ba/Ag and P atoms as shown in the Fig. 4. Hence covalent bond is present among the constitutional atoms of BaAgP. To explore the bonding nature (ionic, covalent or metallic) in the molecule in greater depth, the Mulliken atomic populations [56] and Mulliken charge of BaAgP have been evaluated and are listed in Table 4. Mulliken charge analysis shows that the atomic charge assigned to Ba, Ag and P are 0.97, 0.16 and -1.13 electronic charge, respectively. This indicates that charge is transferred from Ba and Ag atoms to P atom. The transfer of charge between the constitutional atoms of BaAgP indicates the presence of ionic nature in their chemical bonds. Thus the bonding in BaAgP is expected to be a mixture of covalent and ionic.

**Table 4**. Mulliken atomic population and Mulliken charge of BaAgP.

| Compound | Atom | Mulliken atomic population | | | | | |
| | | *s* | *p* | *d* | *f* | Total | Charge (e) |
| --- | --- | --- | --- | --- | --- | --- | --- |
| | Ba | 2.00 | 6.05 | 0.98 | 0.00 | 9.03 | 0.97 |
| BaAgP | Ag | 0.86 | 0.13 | 9.84 | 0.00 | 10.84 | 0.16 |
| | P | 1.90 | 4.24 | 0.00 | 0.00 | 6.13 | -1.13 |

### 3.3 Phonon Properties

The phonon dispersion spectra (PDS) of a material yields valuable information regarding structural stability, phase transition and vibrational contribution in properties such as thermal expansion, Helmholtz free energy, and heat capacity [57]. The phonon dispersion spectra of BaAgP along the high symmetry direction of the Brillouin zone (BZ) are investigated. The density functional perturbation theory (DFPT) based linear response method [58-60] is employed to calculate these properties which are shown in Fig. 5. The PDS of BaAgP exhibit positive phonon frequencies and no region of negative frequencies in the whole region depicted. As there are no imaginary frequency branch exists in the whole Brillouin zone, BaAgP is dynamically stable at zero pressure. The lower branches in the dispersion spectra are the acoustic branches, which originate as a result of coherent movements of atoms of the lattice outside of their equilibrium positions. The acoustic modes at the $\Gamma$ point have zero frequency. This is also a sign of dynamical stability of the studied compound [61]. Again, the upper branches represent the optical branches which have non-zero frequency at the $\Gamma$ point. The optical properties of crystals are mainly controlled by the optical branches [61]. Moreover, there is no phonon gap between acoustic and optical branches, indicating a strong optical-acoustic phonon scattering which will suppress the lattice thermal conductivity, $\kappa_l$ [62] of material. We have also calculated the total and partial density of phonon states (the right panel of Fig. 5). The PHDOS can be divided into three regions: acoustic modes, lower optical modes and upper optical modes. It is seen from the phonon DOS curve that Ba atom mainly contribute to the acoustic phonon modes. Lower optical modes contain the contribution of all the atoms. But only P atom contribute to the upper optical modes. This is expected because P is lighter than other atoms.

### 3.4 Optical properties

In order to describe the response of BaAgP to electromagnetic radiation, we have calculated several optical parameters. The response to the IR, visible and UV spectra is important from the view point of optoelectronic applications. We have calculated all the parameters using TB-mBJ functional. The GGA-PBE functional has been neglected due to its underestimation of energy band gap. The calculated optical parameters of BaAgP for photon energies up to 14 eV have been presented in Fig. 6. All the parameters have been calculated along $a$ (parallel) direction and $c$ (perpendicular) direction due to the hexagonal structure of BaAgP. The complex dielectric function, $\varepsilon(\omega) = \varepsilon_1(\omega) + i\varepsilon_2(\omega)$ (where $\varepsilon_1(\omega)$ is the real and $\varepsilon_2(\omega)$ is the imaginary part), shows the optical response of a medium at all photon energies. For metallic materials both intraband and interband transitions contribute to the dielectric function. However, intraband transitions have very small contributions to the dielectric function for a semiconductor. Hence, in our present calculations we have considered only interband transitions. The interband transition is further divided into the direct band and the indirect band transitions. The indirect interband transition can be neglected as it has a little contribution to $\varepsilon(\omega)$, which involves electron phonon scattering [63]. The calculated real and imaginary part of the dielectric function of BaAgP is presented in Fig. 6 (a, b). The static dielectric constant at zero frequency, $\varepsilon_1(0)$ of BaAgP is 10.37 along $a$-axis and 7.37 along $c$-axis. This value is inversely related to the band gap. It is somewhat smaller than that of well-known photovoltaic materials silicon (11.7) and GaAs (12.9). Beyond this limit, $\varepsilon_1(\omega)$ increases and reaches a maximum value of 17.46 at 2.05 eV along $a$-axis and 13.06 at 2.53 eV along $c$-axis. After reaching the maximum value, $\varepsilon_1(\omega)$ starts to decrease. Both parallel and perpendicular components of $\varepsilon_1(\omega)$ goes below 0.0 at an energy of 4.95 eV. For negative values of $\varepsilon_1(\omega)$, the material lost the dielectric property [64]. The maximum peak of $\varepsilon_2(\omega)$ is found to be in the visible region at 2.40 eV and it is 15.97 along $a$-direction. On the other hand, along $c$-direction, $\varepsilon_2(\omega)$ has two peaks, one is at 3.09 eV (10.75) and another is at 4.20 eV (11.16). The variation of refractive index as a function of photon energy is plotted in Fig. 6(c). Interestingly, it follows the pattern of $\varepsilon_1(\omega)$. The static value of refractive index, $n(0)$ is 3.22 along $a$-axis and 2.72 along $c$-axis. This value is close to the value of GaAs (3.29 -3.86) [65, 66] and silicon (3.88) [67]. Therefore, BaAgP is a better candidate for photovoltaic material. The static refractive index, $n(0)$ satisfies the relation $n(0)^2 \approx \varepsilon_1(0)$, which verifies the accuracy of the calculation [68]. It is observed from the Fig. 6(c) that beyond $n(0)$, refractive index starts to increase up to ~2.10 eV and reaches a maximum value of 4.33 along a-axis and along c-axis this value is 3.70 at ~2.62 eV. We observed that $n(0)$ lies in the infrared region, which increases to a maximum value in the visible region and then it starts decreasing and falls below unity at ~8 eV. The material is transparent if $n(\omega)$ is around zero whereas the positive values measure the absorption of light. Fig. 6(d) shows the photon energy (frequency) dependent extinction coefficient for BaAgP. It is observed that the trends of $k(\omega)$ is similar to that of $\varepsilon_2(\omega)$ because it also provides a measure of absorption of incident radiation [68]. The maxima of extinction coefficient occurs at about 2.63 eV along $a$-axis and 3.18 eV along $c$-axis. The extinction coefficient is very low up to ~1.65 eV. It indicates very low absorption of light which is favorable for transparent properties of the material in this range of the photon energy [68]. The reflectivity $R(\omega)$ vs energy curve of BaAgP semiconductor is shown in Fig. 6(e). The figure shows that the reflectivity remains almost constant from visible to ultraviolet region (1.65 - 12 eV). Reflectivity is ~ 46% along $a$- axis and ~37% along $c$-axis in the visible region. The static value of reflectivity, $R(0)$ is 0.3 ($a$-axis) and 0.2 ($c$-axis). Our calculated result for the absorption coefficient of BaAgP is correspondingly shown in Fig. 6(f). Various peaks are seen in the absorption spectra of the compound. Such peaks are due to the interband transition from bonding band to the anti-bonding band, respectively. The maximum value of $\alpha(\omega)$ is around $127 \times 10^4 \, \text{cm}^{-1}$ (along $a$-axis) at ~9 eV and $138 \times 10^4 \, \text{cm}^{-1}$ (along $c$-axis) at ~11 eV. It is observed that $\alpha(\omega)$ remains almost constant in the energy range 6 – 14 eV. The value of absorption coefficient of BaAgP is much larger than that of GaAs ($(14\text{-}22) \times 10^4 \, \text{cm}^{-1}$ at ~4.8 eV) [69, 70] but close to the value for $Ba_3SbN$ and $Ba_3BiN$ ($(120 \times 10^4 \, \text{cm}^{-1}$ at ~8 eV) [71] and smaller than that of the value for silicon ($180 \times 10^4 \, \text{cm}^{-1}$) [72]. In the visible region, the maximum value of the parallel component of absorption coefficient is $64 \times 10^4 \, \text{cm}^{-1}$ and the perpendicular component is $54 \times 10^4$ cm$^{-1}$. Such a high absorption of incoming radiation makes this compound favorable for applications in photovoltaic devices. Fig. 6(g) shows that the real part of optical conductivity does not start at zero photon energy which indicates that an energy band gap exists in the compound. Hence this compound is a narrow band gap semiconductor which is evident from our band structure plot (Fig. 2). Furthermore, the optical conductivity and hence electrical conductivity of the material increases as a result of photon absorption. The maximum value of optical conductivity, in the visible region, is of $5.4 \times 10^3$ (ohm-cm)$^{-1}$ along *a*-axis and $4.4 \times 10^3$ (ohm-cm)$^{-1}$ along *c*-axis. Energy-loss function is an important optical parameter which describes possible interactions of a material by fast moving electrons. Such interactions are responsible not only for interband transition but also a source of interband transitions, inner shell ionizations, phonon excitation, and plasmon excitations. Fig. 6(h) shows energy loss spectra $L(\omega)$ for BaAgP plotted in the energy range 0–14 eV. It is seen from the figure that the prominent peaks are located at 11.82 eV (*a*-axis) and 11.97 eV (*c*-axis) for BaAgP. These peaks correspond to the rapid decrease of reflectivity spectra $R(\omega)$. The analysis of the calculated optical parameters suggests that BaAgP is optically anisotropic and a potential candidate for optoelectronic device applications. Unfortunately we could not compare our optical results as there are no experimental as well as theoretical studies.

### 3.4 Thermoelectric transport properties

The performance of a thermoelectric material depends on its Seebeck coefficient, electrical conductivity and thermal conductivity. To realize the thermoelectric response and applicability of the studied compound we have calculated all these parameters using the Boltzmann semi-classical transport theory as implemented in BoltzTrap code. Fig. 7 shows the variation of thermoelectric parameters of BaAgP as a function of temperature. The parallel (along *a*-axis) and perpendicular components (along *c*-axis) of these parameters are shown with blue and red curves, respectively. We have calculated the thermoelectric parameters considering TB-mBJ functional only as the PBE functional underestimate the band gap. It is noted from Fig. 7(a) that Seebeck coefficient, $S$ shows completely different behavior along $a$- and $c$-axes. The parallel component of the Seebeck coefficient sharply decreases with increasing temperature up to 300K. Beyond this range of temperature it remains almost constant up to 1000K. Again, the magnitude of the perpendicular component of $S$ increases with temperature up to 200K and then it gradually decreases. At room temperature, the value of the parallel and perpendicular components of Seebeck coefficient are 240 µV/K and 236 µV/K, respectively. The positive value of the Seebeck coefficient indicates that BaAgP is a $p$-type semiconductor. The electrical conductivity scaled by the relaxation time ($\sigma/\tau$) as a function of temperature is shown in Fig. 7(b). It is observed that both parallel and perpendicular components of electrical conductivity increase with temperature but the rate of increase of parallel component is larger than that of perpendicular component. The electronic thermal conductivity scaled by the relaxation time $\kappa_e/\tau$ (Fig. 7(c)) slowly increases with temperature up to 700 K and then it exhibits similar character like electrical conductivity. These is because the electrons are thermally excited with the increase of temperature and conducted more heat. A good thermoelectric material requires a large value of power factor as well as minimum thermal conductivity. The value of power factor ($S^2\sigma$) depends on both the Seebeck coefficient and electrical conductivity of a material. Fig. 7(d) shows that power factor increases with temperature and exhibits analogues behavior like electrical conductivity. We have also studied the Seebeck coefficient, electrical conductivity, electronic thermal conductivity and power factor as a function of carrier concentration for two fixed values of temperature 700K and 1000K and shown in Fig. 8. Seebeck coefficient for bulk semiconducting material can be expressed as $S = \frac{8\pi^2 k_B^2 T}{3eh^2} m^* \left(\frac{\pi}{3n}\right)^{2/3}$ [73]. Hence $S$ largely depends on carrier concentration, carrier effective mass and density of states (DOS). Fig. 8 (a) shows that Seebeck coefficient first increases at low carrier concentration ($n < 10^{19}$ cm$^{-3}$ for 700K and $n = 10^{19}$ cm$^{-3}$ for 1000K), reaches its peak and then decreases at high carrier concentration ($10^{19} < n > 10^{22}$ cm$^{-3}$) for both 700 and 1000K. This is mainly due to the bipolar effect resulting from a strong thermal excitation of minor carriers [74]. For the low carrier concentration ($n \leq 10^{19}$ cm$^{-3}$), $S$ is regulated by the carrier effective mass $m^*$ and the magnitude of DOS around the Fermi level. A smoother valence band (hole-doping) compared with a conduction band (electron-doping) corresponds to a larger carrier $m*$, indicating a large $S$ since $S$ is proportional to $m^*$ [73]. On the other hand, for high carrier concentration ($10^{19} < n > 10^{22}$ cm$^{-3}$), $S$ decreases with the increasing carrier density $n$ because $S$ is inversely related to $n$. The value of electrical conductivity scaled by the relaxation time, $\sigma/\tau$ is very small up to the carrier concentration $10^{19}$ cm$^{-3}$ because $\sigma$ is inversely related to $m^*$. After then it gradually increases with the increase of carrier concentration for both 700 and 1000K temperature. $\kappa/\tau$ shows similar character but in this case its value is slightly larger for 1000K than that for 700K. This is due to the excitation of charge carriers at higher temperature. Fig. 8(d) shows the variation of power factor (PF) with charge carrier concentration for 700K and 1000K. The PF exhibits different trend along $a$- and $c$-direction. The parallel component of PF first increases with the increasing carrier concentration and after reaching the maximum value it starts to decrease. The maximum value of PF (along $a$-axis) is obtained for the carrier density of $1.0 \times 10^{20}$ cm$^{-3}$ at 700K and $2.5 \times 10^{21}$ cm$^{-3}$ at 1000K. The perpendicular component of PF is maximum for the carrier density of $10^{22}$ cm$^{-3}$ at both 700K and 1000K. The temperature dependent lattice thermal conductivity, $\kappa_L$, thermal conductivity, $\kappa$ and figure of merit, ZT for BaAgP are depicted in Fig. 9. The lattice thermal conductivity is one of the important physical properties of a compound, especially for applications at elevated temperatures. The empirical formula for lattice thermal conductivity due to Slack [75-77] is $\kappa_l = A(\gamma) \frac{M_{Av}\theta_D^3 \delta}{\gamma^2 n^{2/3}T}$ , where $M_{av}$ denotes the average atomic mass (in kg/mol) in a crystal, $\delta$ is the cubic root of average atomic volume, $n$ is the total number of atoms in the unit cell, $T$ is the absolute temperature, $\gamma$ is the Grüneisen parameter and $A$ is a $\gamma$ dependent factor (in W-mol/kg/m$^2$/K$^3$). The parameter $\gamma$ can be calculated using Poisson's ratio $\nu$ from, $\gamma = 1.5(1+\nu)/(2-3\nu)$. The formula for $A(\gamma)$ involving $\gamma$ is given by Julian [78]. The Grüneisen parameter can partially address the issue of anharmonicity of phonon–phonon interaction [79]. The temperature dependent lattice thermal conductivity is depicted in Fig. 9 (a). This shows a $T^{-1}$ behavior of $k_L$ with temperature, which is due to the Umklapp phonon–phonon scattering process [80]. According to the Wiedemann–Franz law, the thermal conductivity ($\kappa = \kappa_e + \kappa_l$) is the sum of electronic thermal conductivity ($\kappa_e = L\sigma T = ne\mu LT$) and lattice thermal conductivity ($\kappa_L$). Here $L$ is the Lorenz factor, $2.4 \times 10^{-8}$ J$^2$ K$^{-2}$ C$^{-2}$ for free electrons. Our calculated total thermal conductivity as a function of temperature is presented in Fig. 9 (b). To calculate total thermal conductivity we have multiplied electronic thermal conductivity by $\tau = 10^{-14}$ s and adding it with lattice thermal conductivity. The figure shows that both the components (a- and c-axes) of total thermal conductivity decreases with the increase of temperature. It can be noted from the figure that the contribution of $\kappa_l$ is larger than that of $\kappa_e$ to the total thermal conductivity. In semiconductors, the heat energy is mainly conducted by phonons due to its negligible conduction electron density and hence $\kappa_l$ dominates. The lowest value of thermal conductivity of a perfect crystal at high temperature is the intrinsic minimum thermal conductivity when phonons completely uncouple and hence the thermal conductivity reaches to the minimum value [81]. In this condition, the mean free path of phonons may be approximated as the average interatomic distance. In this case total mass $M$ in the unit cell may be replaced by an equivalent atom with an average atomic mass of $M/n$ [82]. A single equivalent atom in the cell can be used to obtain a formula for $\kappa_{\min}$ [83] at high temperature given by $\kappa_{min} = k_B v_m \left(\frac{M}{n \rho N_A}\right)^{-2/3}$. For BaAgP, our calculated value of minimum thermal conductivity is 0.513 W/mK. Therefore, it is clear that the lattice thermal conductivity strongly depends on phonon mean free path. If we can reduce phonon mean free path which is comparable with interatomic distance, the minimum thermal conductivity could be achieved. The nano-structuring can be the possible route for the reduction of lattice thermal conductivity because in this case phonon scattering increases. The thermoelectric performance of a material is characterized by its dimensionless figure of merit $ZT\left(=\frac{S^2\sigma}{\kappa}T\right)$. The temperature dependence ZT for BaAgP is depicted in Fig. 9(c). It is noted that the parallel component of ZT increases sharply with increasing temperature than that of its perpendicular component. This is due to larger power factor along a-direction. For BaAgP the thermoelectric figure of merit at 1000 K is ~ 0.44. Hence BaAgP is a potential thermoelectric material. The comparison of the values of parallel and perpendicular components of the transport parameters show that transport properties of BaAgP is anisotropic. This is due to the large variation of lattice parameters for both components. To the best of our knowledge, the thermoelectric properties of BaAgP is calculated for the first time. Hence we could not compare our outcomes to experimental or other theoretical results.

## 4. Conclusions

In this report, we have explored the mechanical, electronic, optical, phonon and thermoelectric properties of *p*-type half-Heusler compound BaAgP for the first time using DFT-based first principles calculations. Our calculated lattice parameters agree very well with the experimental values. The analysis of mechanical and phonon properties reveal that BaAgP is mechanically and dynamically stable. This compound is brittle, soft and elastically anisotropic in nature. The linear compressibility along *c*-axis is almost double than that of along *a*-axis. The analysis of electronic band structure shows that this compound has a narrow indirect band gap and its value is 0.59 for TB-mBJ functional. The charge density distribution map and Mulliken bond population show that this compound contains a mixture of covalent and ionic bonding. The study of optical properties of BaAgP show that this compound is optically anisotropic. It has high absorption coefficient and low reflectivity in the visible to ultraviolet region which shows that this compound could be a possible candidate of solar cell and optoelectronic device applications.  The calculated power factor at 1000K along a-axis is 35.2 µW/cmK$^2$ (with τ=10$^{-14}$ s) which is ~3.5 times larger than that (10.1 µW/cmK$^2$ along b-axis at 850 K) of SnSe, a promising layered thermoelectric materials [84]. A layered material exhibit low thermal conductivity because phonon scattered between two layers in a particular direction. The total thermal conductivity, κ (with τ=10$^{-14}$ s) along a-axis at 1000 K  is ~ 8 W/mK, which is much higher than that (0.34 W/mK at 900 K along b-axis) of SnSe [84]. The calculated ZT value is 0.44 and the large thermal conductivity is responsible for this low value. Therefore, the reduction of thermal conductivity of BaAgP can be achieved by applying suitable techniques (Nano-structuring or band engineering) and hence the higher ZT value could be achieved. The obtained results of this investigation will inspire experimentalists to carry out experiment for the reduction of thermal conductivity and check thermoelectric performance of BaAgP.

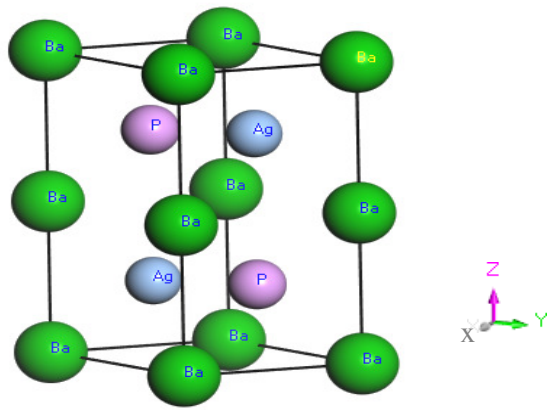

**Fig. 1.** The unit cell of BaAgP.

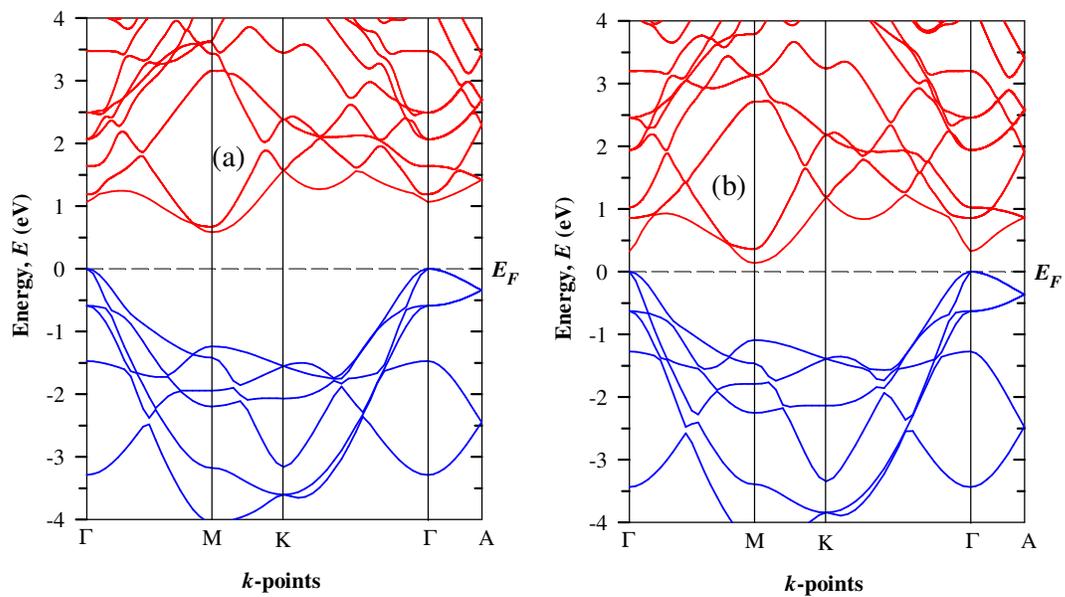

**Fig. 2.** Electronic band structure along high symmetry directions of the Brillouin zone of BaAgP for (a) TB-mBJ and (b) PBE functional.

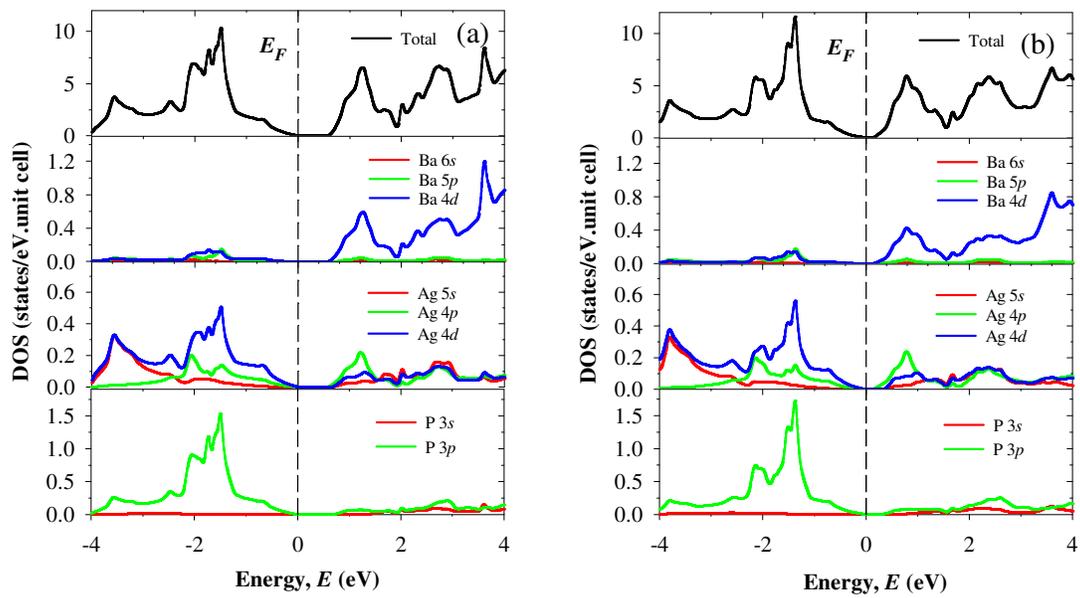

**Fig. 3**. The total and partial electronic density of states of BaAgP for (a) TB-mBJ and (b) PBE functional.

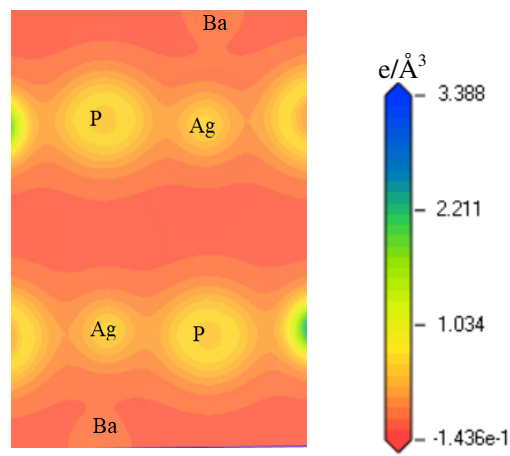

**Fig. 4.** Electronic charge density map of BaAgP.

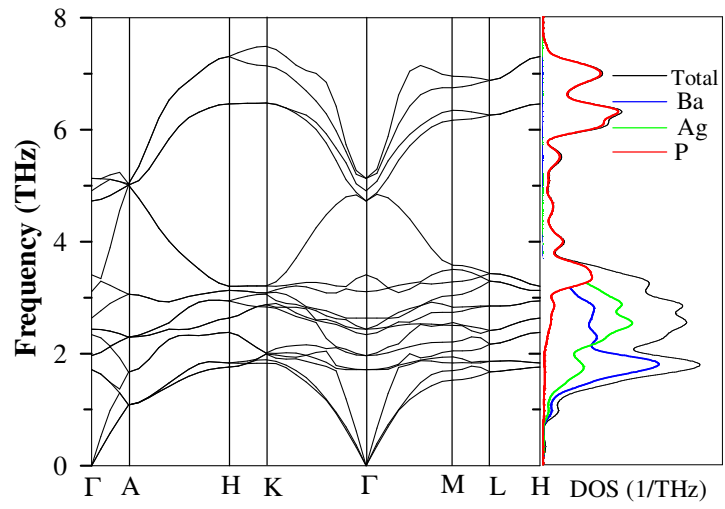

**Fig. 5.** Calculated phonon dispersion spectra and phonon DOS for BaAgP.

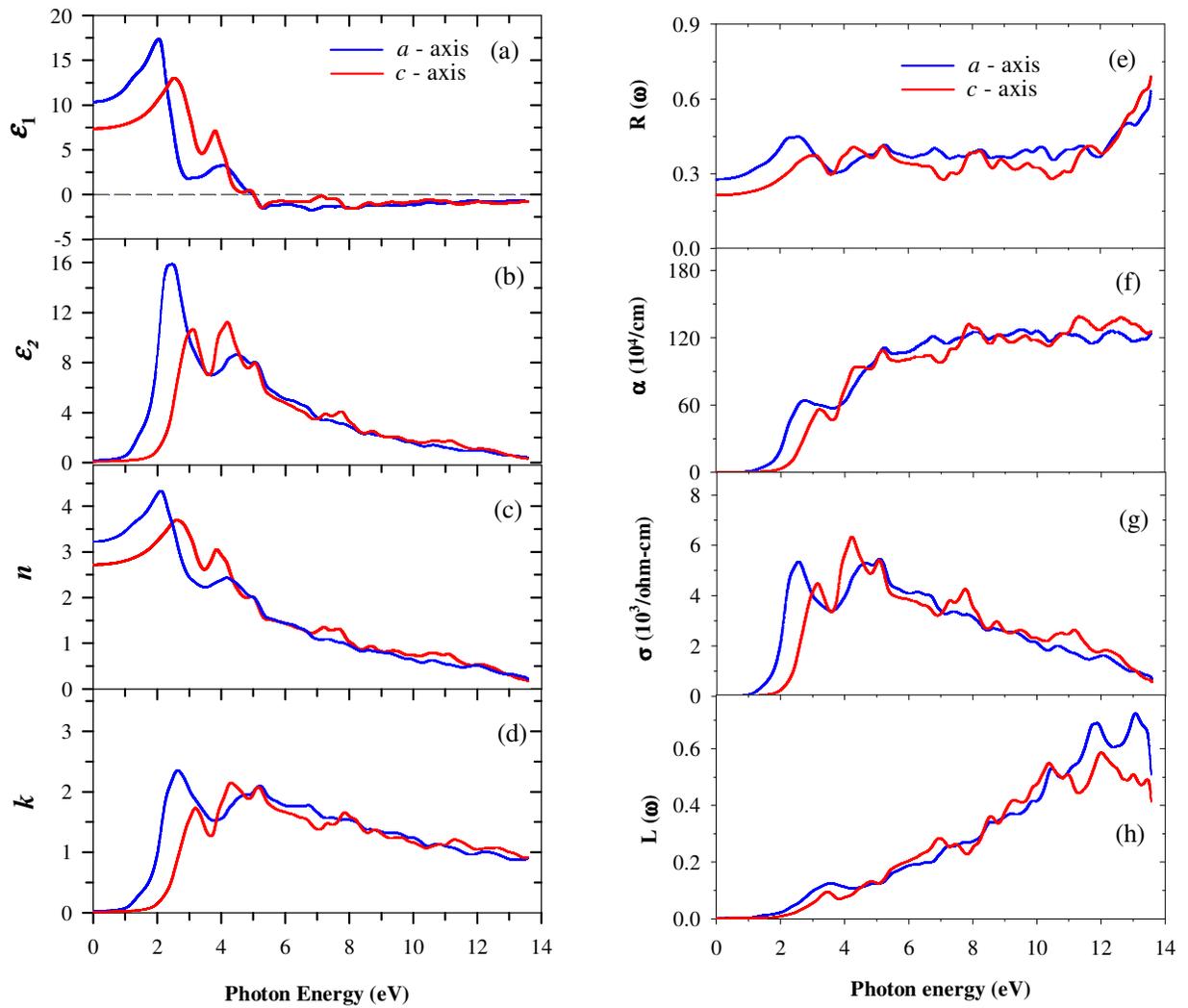

**Fig. 6.** Photon energy dependence of (a) real, $\varepsilon_1$ (b) imaginary, $\varepsilon_2$ part of dielectric function (c) refractive index, $n$ (d) extinction coefficient, $k$ (e) reflectivity, $R$ (f) absorption, $\alpha$ (g) optical conductivity, $\sigma$ and (h) loss function, L of BaAgP.

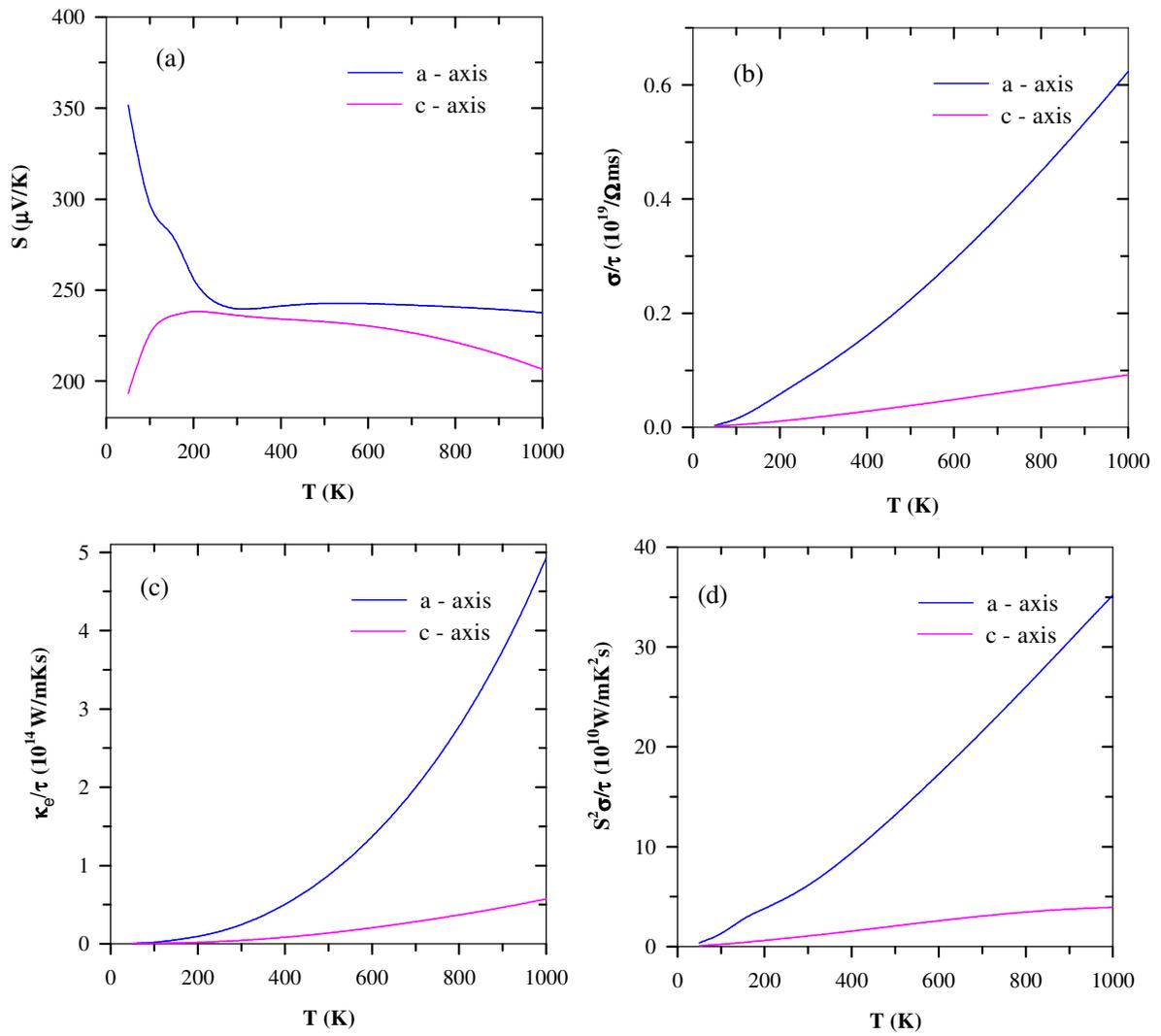

**Fig. 7**. Temperature dependent thermoelectric transport properties of BaAgP: (a) Seebeck coefficient (b) electrical conductivity (c) electronic part of the thermal conductivity and (d) Power factor.

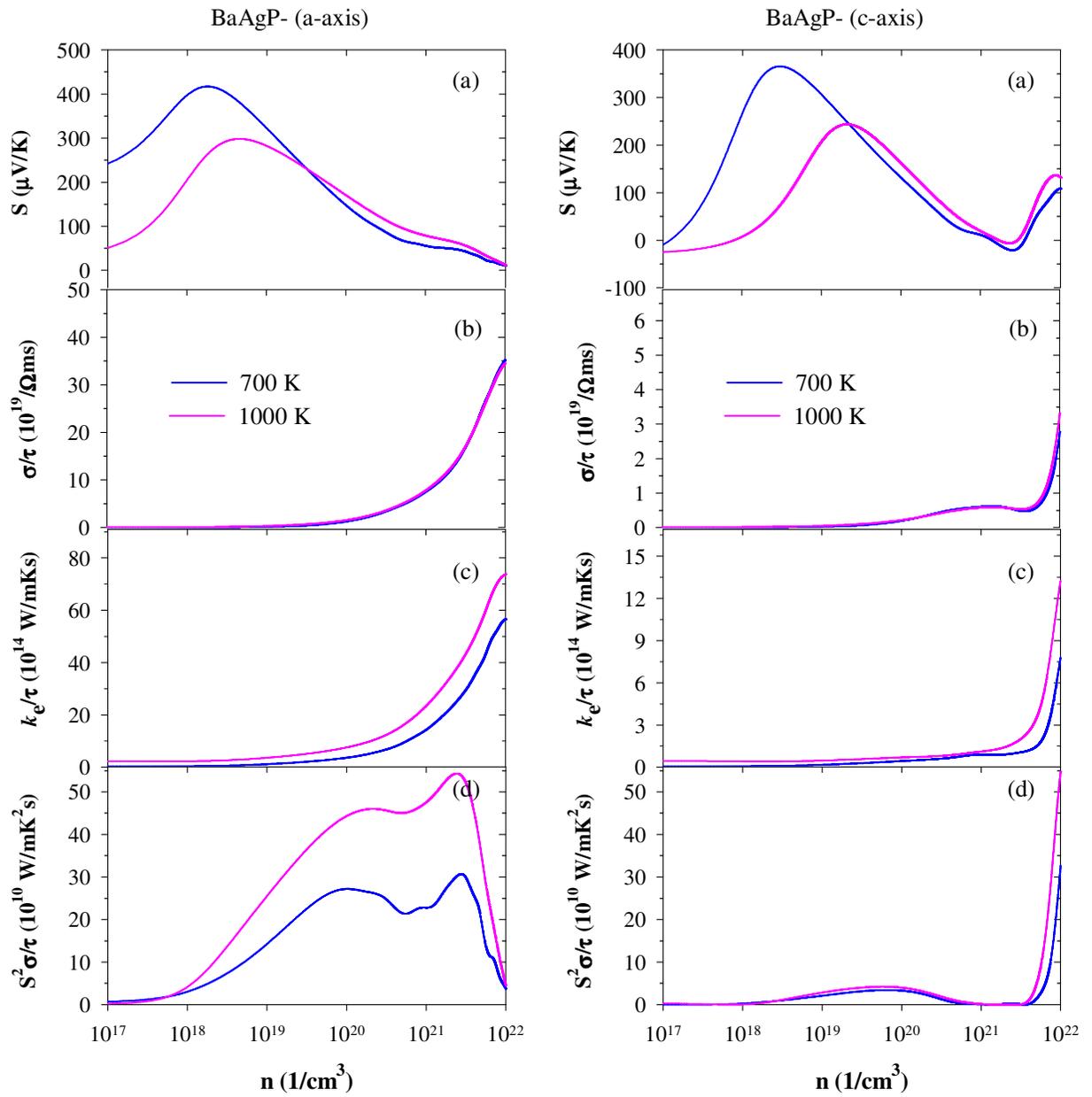

**Fig. 8.** Thermoelectric parameters as a function of carrier concentration. (a) Seebeck coefficient, (b) electrical conductivity, (c) electronic thermal conductivity, (d) Power factor. The left panel shows the parallel (along *a*- axis) component and right panel shows the perpendicular (along *c*-axis) component of the parameters.

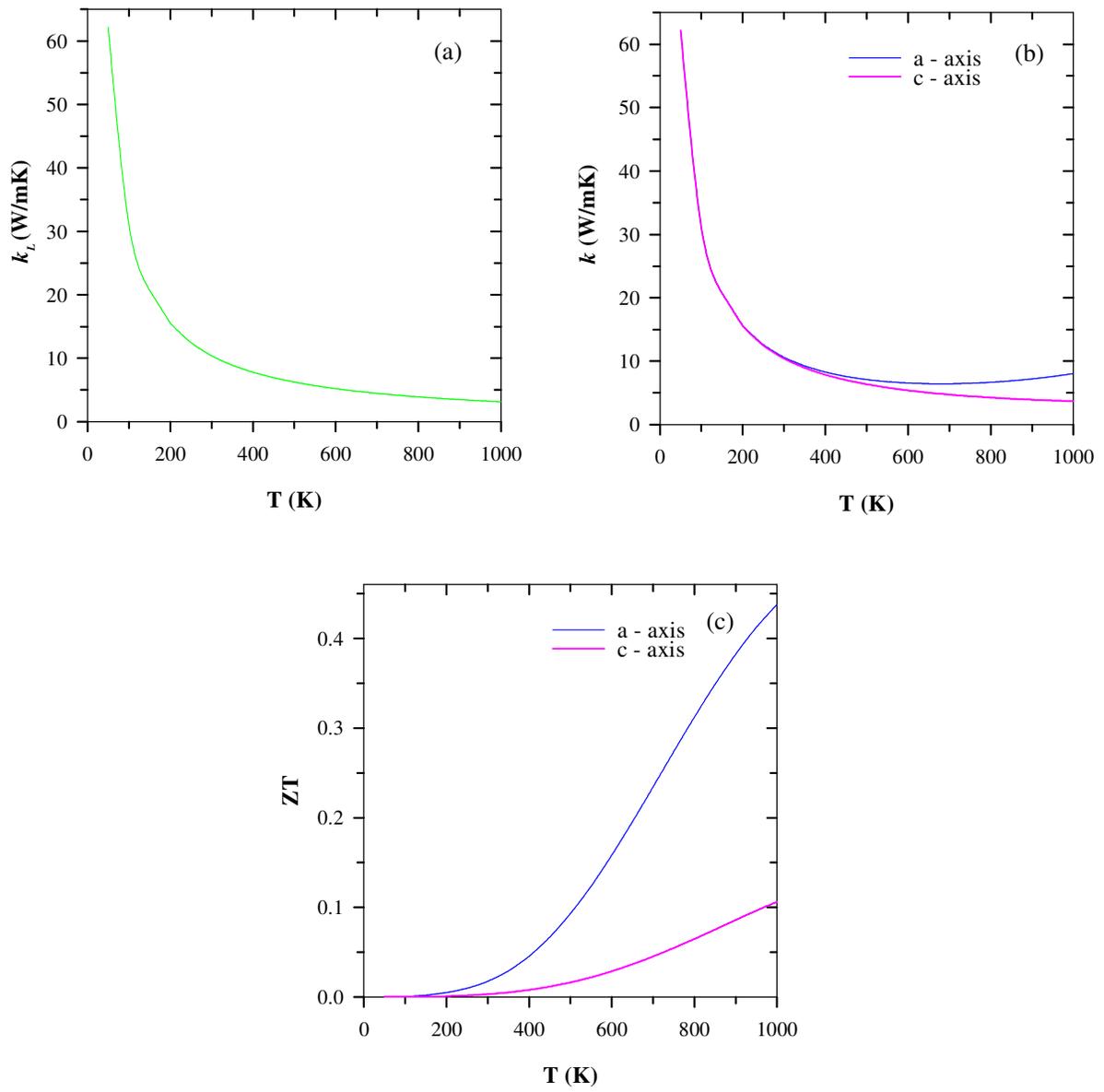

**Fig. 9.** Temperature dependence of (a) lattice thermal conductivity, $\kappa_L$, (b) total thermal conductivity, $\kappa$ and (c) figure of merit ZT of BaAgP.